\documentclass[10pt,letterpaper]{article}

\usepackage{amapvoice}

\reporttitle{PilotTTS}
\headerinst{Alibaba Group,\ 2026}
\hypersetup{pdftitle={PilotTTS: A Disciplined Modular Recipe for Competitive Speech Synthesis}}

\begin{document}
\thispagestyle{firstpage}

\begin{center}

{\small\color{TextGray}%
  \textsc{Amap Voice}~$\cdot$~Technical Report~$\cdot$~2026}

\vspace{10pt}

{\LARGE\bfseries PilotTTS: A Disciplined Modular Recipe for Competitive Speech Synthesis}

\vspace{13pt}

{\normalsize\bfseries
  Bowen Li$^{1,*}$\enspace
  Shaotong Guo$^{1,*}$\enspace
  Zhen Wang$^{1,*, \dagger}$\enspace
  Yang Xiang$^{1}$\enspace
  Mingli Jin$^{1}$\enspace
  Yihang Lin$^{1,2}$\enspace
  Jiahui Zhao$^{1}$\\[2pt]
  Weibo Xiong$^{1}$\enspace
  Dongrui Zhang$^{1}$\enspace
  Keming Chen$^{1}$\enspace
  Yunze Gao$^{1}$\enspace
  Zeyang Lin$^{1, \ddagger}$\enspace
  Yuze Zhou$^{1}$\enspace
  Yue Liu$^{1}$\enspace
}

\vspace{6pt}

{\small\color{TextGray}
  $^{1}$Amap, Alibaba Group\quad
  $^{2}$The Chinese University of Hong Kong, Shenzhen
}

\vspace{4pt}

\renewcommand{\thefootnote}{}
\footnotetext{%
  \begin{tabular}[t]{@{}l@{~}l}
    $*$ & Equal contribution.\\
    $\dagger$ & Project leader.\\
    $\ddagger$ & Corresponding author: \texttt{linzeyang.lzy@alibaba-inc.com}
  \end{tabular}%
}
\renewcommand{\thefootnote}{\arabic{footnote}}

\vspace{14pt}
\end{center}

\begin{abstractbox}
{\small
Building state-of-the-art text-to-speech (TTS) systems typically demands millions of hours of proprietary data and complex multi-stage architectures, creating substantial barriers for resource-constrained research teams. In this report, we present PilotTTS, a lightweight autoregressive TTS system that achieves competitive performance through minimalist architecture and rigorous data engineering. PilotTTS is trained on only 200K hours of data processed entirely with open-source tools. Specifically, our contributions are: (1)~a reproducible multi-stage data processing pipeline covering quality assessment, label annotation, and filtering, and (2)~a compact model architecture that employs Q-Former-based conditioning to decouple speaker identity from speaking style via cross-sample paired training. Within a unified framework, PilotTTS supports zero-shot voice cloning, emotion synthesis (11 categories), paralinguistic synthesis (4 categories), and Chinese dialect synthesis (14 dialects). On the Seed-TTS Eval benchmark, PilotTTS achieves the lowest WER of 1.50\% on \textit{test-en}, a CER of 0.87\% on \textit{test-zh}, and the highest speaker similarity on both test sets (0.862 and 0.815), outperforming systems trained on significantly larger datasets. We release the complete data pipeline recipe, pretrained weights, and code at   \url{https://github.com/AMAPVOICE/PilotTTS}.
}
\end{abstractbox}

\vspace{10pt}

\section{Introduction}

\begin{figure}[h!]
    \centering
    \includegraphics[width=0.8\linewidth]{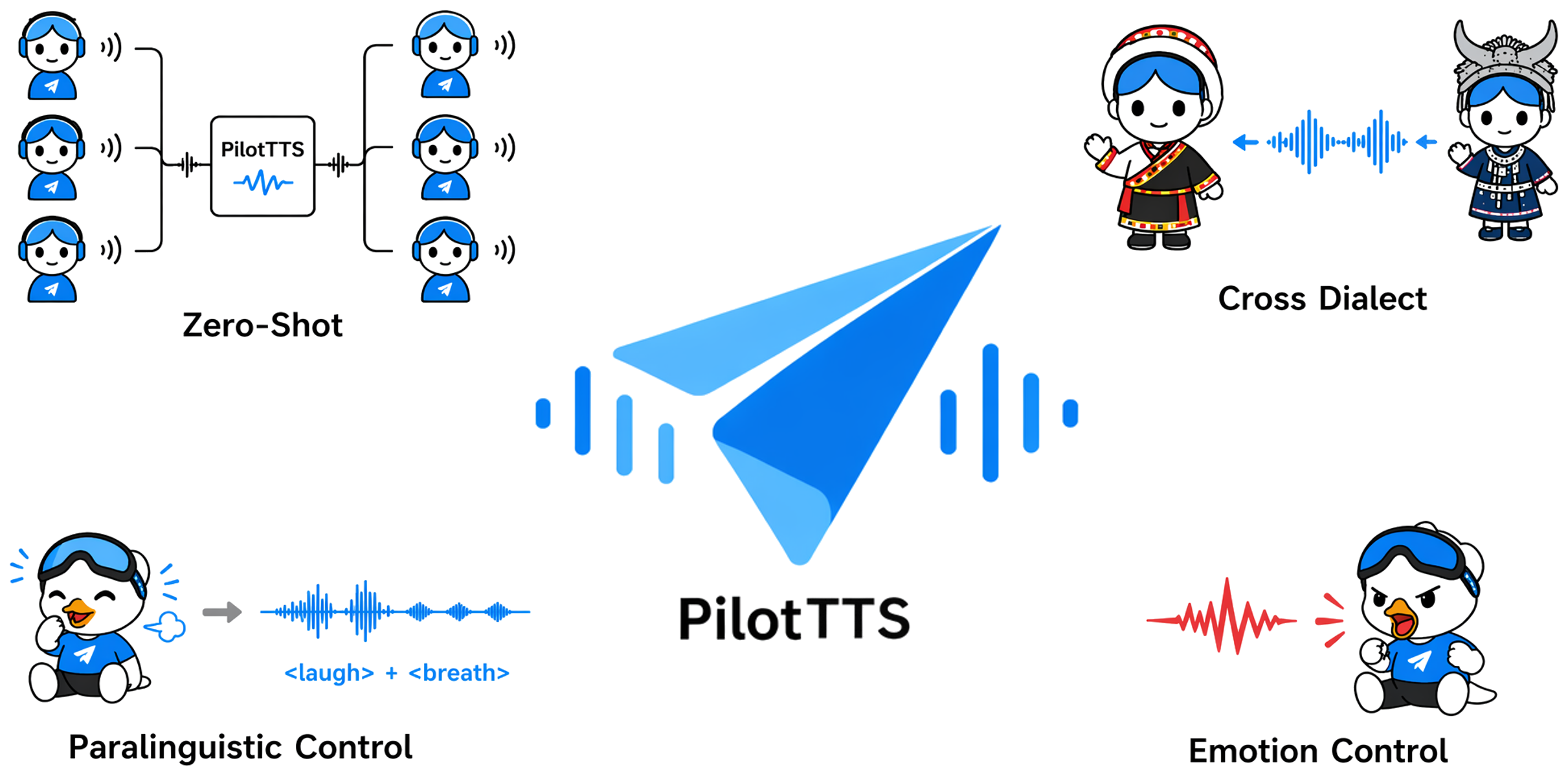}
    \caption{Overview of PilotTTS.}
    \label{fig:introduction}
\end{figure}

With the rapid advancement of generative neural networks, text-to-speech (TTS) synthesis has achieved remarkable progress in naturalness, expressiveness, and speaker fidelity~\cite{chen2025neural,shen2024naturalspeech2,seed-tts}. A particularly transformative development is large-scale zero-shot TTS, where models trained on extensive multi-speaker corpora can clone arbitrary voices from only a few seconds of reference audio. Recent systems built upon large language model (LLM) architectures~\cite{cosyvoice,cosyvoice2,seed-tts,fireredtts2} or non-autoregressive flow-based generation~\cite{f5tts,ditar,naturalspeech3} have demonstrated impressive quality, establishing new standards for the field. However, this progress has been accompanied by a persistent trend toward ever-larger datasets (from hundreds of thousands to millions of hours~\cite{qwen3tts,minimax-speech}) and increasingly complex architectures involving multi-codebook tokenizers, multi-stage training pipelines, and multiple specialized sub-models.

This trajectory, while yielding state-of-the-art results, poses significant challenges for the broader research community. First, acquiring and processing millions of hours of speech data requires proprietary infrastructure and substantial computational resources, with few standardized open-source tools available for constructing high-quality training sets. Second, architectural complexity, such as multi-codebook residual quantization, hierarchical prediction modules, and streaming-specific designs, increases engineering effort and deployment difficulty. Third, advanced controllability features (emotion synthesis, paralinguistic synthesis, dialect synthesis) are typically developed as separate specialized systems, further fragmenting the required expertise and resources. While these directions undoubtedly push the upper bound of speech synthesis expressiveness and represent valid research goals, most commercial deployment scenarios call for a less complex alternative that still meets practical requirements. In such cases, efforts toward extreme expressiveness can paradoxically become an obstacle to rapid deployment and significantly inflate application cost. These barriers collectively restrict resource-constrained teams from building competitive TTS systems or from extending and refining the approaches they already have access to.

In this report, we present PilotTTS, an autoregressive TTS framework rapidly assembled to meet pressing practical demand, achieving competitive performance by combining well-established open-source components with meticulous data engineering, rather than through architectural novelty or data scale. The system employs a Qwen3 language model~\cite{qwen3} as the autoregressive backbone with a Q-Former-based~\cite{blip2} conditioning module for speaker and style representation, and generates speech through a Conditional Flow Matching (CFM)~\cite{flowmatching} decoder with a Diffusion Transformer (DiT)~\cite{dit} backbone followed by a HiFi-GAN~\cite{hifigan} vocoder. Trained on approximately 200K hours of speech data collected from public sources and processed using a pipeline built entirely from publicly available tools, PilotTTS delivers performance competitive with systems trained on an order of magnitude more data.

Our main contributions are as follows:

\begin{enumerate}
    \item \textbf{A reproducible data processing pipeline built upon publicly available modules.} We design a multi-stage pipeline covering quality assessment and enhancement, label annotation, and quality filtering, assembled entirely from publicly available tools. This pipeline transforms raw internet audio into clean, richly annotated training data at substantially reduced cost. Since every constituent module is openly accessible, any team can readily obtain the underlying components and rapidly assemble their own data processing pipeline, lowering the barrier for TTS data preparation.

    \item \textbf{A compact autoregressive TTS architecture with decoupled conditioning.} We propose a Q-Former-based speaker-style conditioning mechanism paired with a cross-sample training strategy that disentangles static speaker identity from dynamic speaking style. This design achieves the highest speaker similarity (0.862) and highly competitive content accuracy (CER 0.87\%) on the Seed-TTS benchmark, using only 200K hours of training data.

    \item \textbf{Multi-dimensional controllability under low-resource constraints.} Within the same framework, PilotTTS demonstrates strong performance in controllable synthesis across dialect, paralinguistic, and emotional dimensions, supporting emotion control across 11 categories, paralinguistic synthesis across 4 categories (laughter, breathing, crying, and coughing), and Chinese dialect synthesis across 14 Chinese dialects, all achieved through targeted post-training.
\end{enumerate}

The remainder of this report is organized as follows. Section~\ref{sec:data} describes our data processing pipeline. Section~\ref{sec:method} details the model architecture and training strategy. Section~\ref{sec:experiments} presents experimental results across multiple evaluation dimensions. Section~\ref{sec:conclusion} concludes with a discussion of limitations and future directions.

\section{Data Processing Pipeline}
\label{sec:data}

\begin{figure*}[ht]
    \centering
    \includegraphics[width=0.9\textwidth]{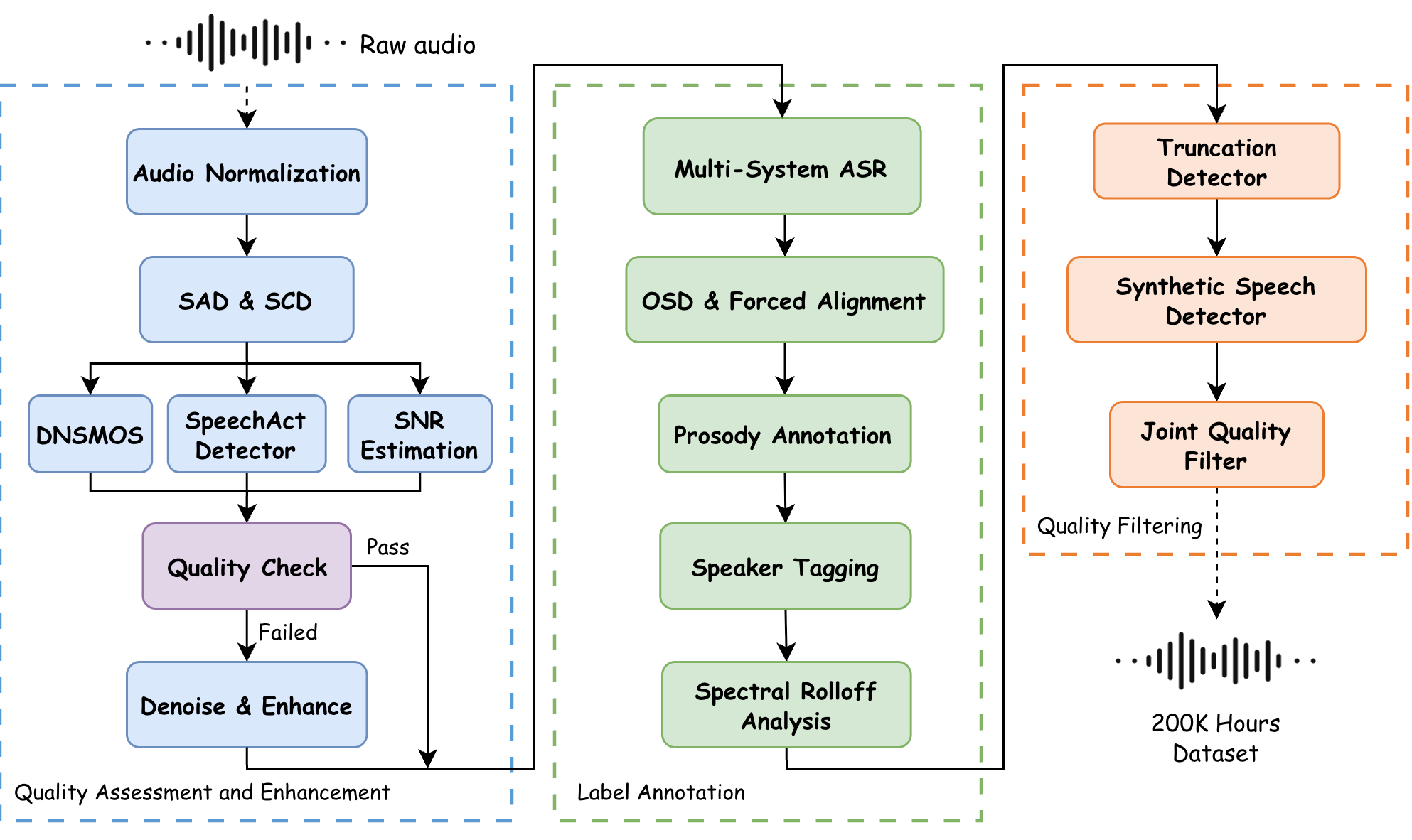}
    \caption{Overview of the three-stage data processing pipeline. Raw audio is first assessed and enhanced (left), then annotated with textual, temporal, and speaker-related labels (middle), and finally filtered through multi-dimensional quality checks to produce approximately 200K hours of training data (right).}
    \label{fig:pipeline}
\end{figure*}

High-quality training data is critical for competitive speech synthesis, yet existing large-scale TTS systems often rely on proprietary data processing pipelines that limit reproducibility. To address this, we design a multi-stage data processing pipeline built entirely upon publicly available modules, composed of three sequential stages: quality assessment and enhancement, label annotation, and quality filtering (Figure~\ref{fig:pipeline}). Starting from raw waveform data collected from heterogeneous real-world sources, the pipeline progressively refines the audio through acoustic preprocessing, structured annotation, and reliability-based sample selection, ultimately producing a clean and richly labeled dataset for downstream model training.

\subsection{Quality Assessment and Enhancement}
\label{sec:quality}

The first stage evaluates and improves the acoustic quality of the input audio. Since raw recordings from diverse sources may exhibit substantial variability in recording conditions, background noise, speaker transitions, and signal quality, this stage combines audio standardization, speech structure analysis, and quality-aware enhancement.

We first standardize all recordings to a unified format and sampling rate to reduce source-level heterogeneity. The standardized audio is then segmented by a speech activity detection (SAD) module together with a speaker change detection (SCD) module to identify valid speech regions and detect speaker transitions~\cite{pyannote-powerset, pyannote}.

Next, three quality-related attributes are estimated in parallel. We use DNSMOS~\cite{dnsmos} to predict perceptual speech quality, a speech/non-speech classifier\footnote{\url{https://www.modelscope.cn/models/iic/SenseVoiceSmall}} to determine whether a segment contains valid speech rather than music, environmental noise, or other non-speech events, and an SNR estimation module to measure the signal-to-noise ratio. Based on these indicators, each segment is assigned a quality status. Segments with low predicted perceptual quality (e.g., predicted MOS~$\leq$~3.5), non-speech labels, or insufficient SNR are regarded as acoustically deficient.

For segments identified as low quality, we apply a denoising and enhancement module\footnote{\url{https://github.com/resemble-ai/resemble-enhance}} to suppress background noise and improve speech intelligibility. This dual assessment-and-enhancement design provides cleaner inputs for subsequent annotation.

\subsection{Label Annotation}
\label{sec:annotation}

The second stage generates structured annotations for each processed speech segment, including textual, temporal, speaker-related, and prosody-related information.

Each audio segment is first transcribed by multiple automatic speech recognition (ASR) systems to obtain textual content. We collect transcription results from several open-source systems, including Paraformer~\cite{paraformer}, FireRedASR~\cite{fireredasr}, and Whisper~\cite{whisper}, as well as internal ASR models. Multi-system transcriptions enable cross-system consistency checking, improving the reliability of the final text labels.

Next, each audio-text pair is processed by an overlapping speech detection (OSD) module\footnote{\url{https://huggingface.co/pyannote/segmentation-3.0}} and a forced alignment module. The OSD component identifies segments containing simultaneous speech from multiple speakers, while the aligner establishes temporal correspondence between the transcribed text and the acoustic signal. Based on the alignment results, we further apply a prosody annotation system built upon Qwen3-Force-Alignment~\cite{qwen3asr} to annotate hierarchical prosodic structures.

We then perform speaker tagging using 3D-Speaker-Toolkit~\cite{3dspeaker} to assign speaker identity metadata to each segment, facilitating the construction of speaker-consistent training samples. In addition, we perform spectral rolloff analysis to identify low-bandwidth recordings with insufficient high-frequency content, preventing audio with excessively low effective sampling rates from entering the training set.

\subsection{Quality Filtering}
\label{sec:filtering}

The final stage removes unreliable or undesirable samples based on the quality signals and annotations accumulated in the previous stages.

We first apply a truncation detector to identify utterances with incomplete beginnings or endings. Such samples are typically caused by inaccurate segmentation or abrupt clipping and can adversely affect model training. We then employ a synthetic speech detector to identify segments that are likely to be artificially generated rather than naturally spoken. This step is particularly important for large-scale crawled data, where synthesized speech may be mixed with authentic recordings.

After these dedicated detection steps, all available metadata and quality indicators are aggregated in a final filtering module. This module performs sample selection by jointly considering acoustic quality, speech validity, transcription reliability, overlap condition, speaker consistency, truncation risk, synthesis likelihood, and spectral quality. Only samples that satisfy all filtering criteria are retained.

Rather than discarding excluded samples, we preserve all processed data together with their quality tags and annotation metadata, enabling the flexible construction of datasets with different quality requirements for future tasks. After filtering, we retain approximately 200,000 hours of Chinese and English speech data for pre-training. All constituent modules used in this pipeline are openly accessible, allowing any team to readily obtain them and rapidly assemble a comparable pipeline tailored to their own data preparation needs.

\section{Method}
\label{sec:method}

\begin{figure*}[ht]
    \centering
    \includegraphics[width=\textwidth]{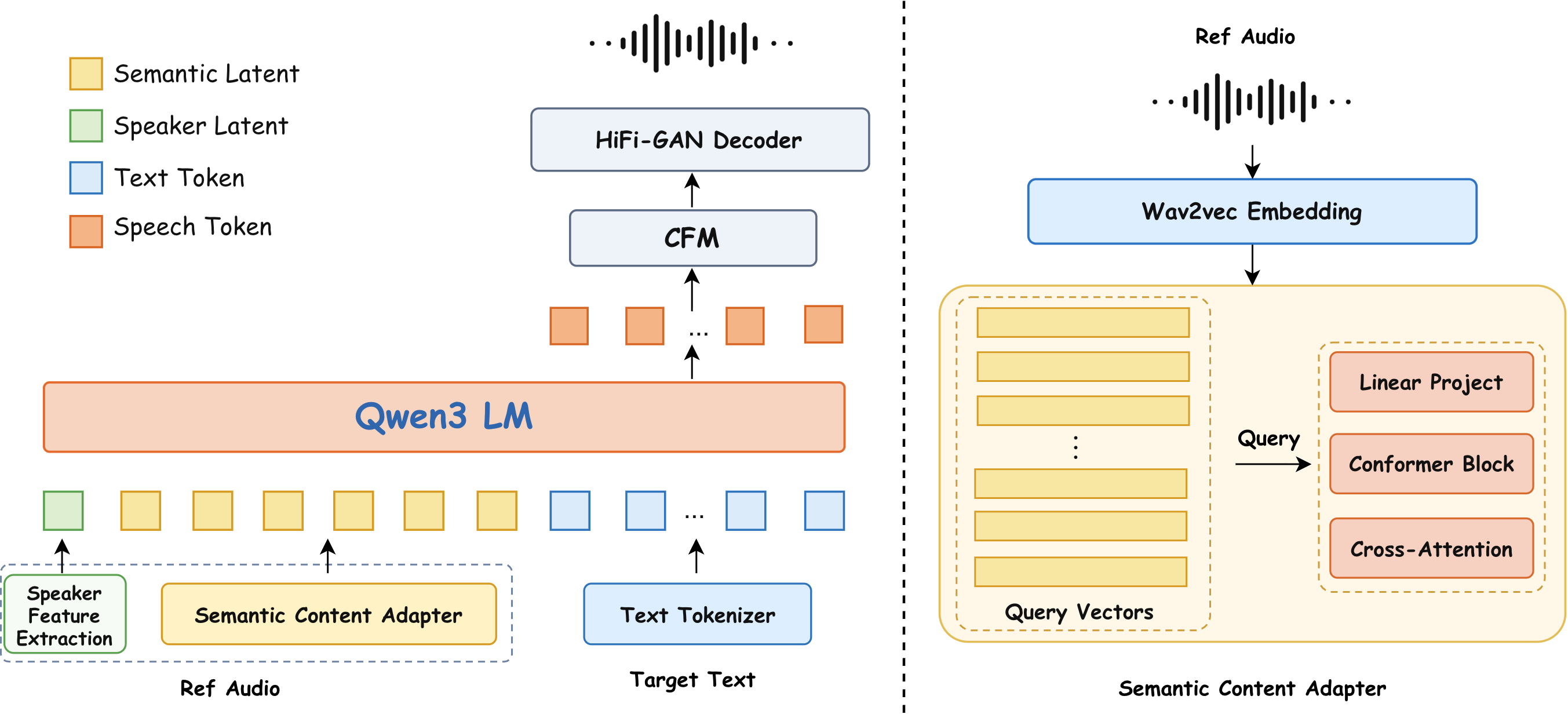}
    \caption{Overall architecture of PilotTTS. Left: the full pipeline comprising a Speaker Feature Extraction module (CAMPPlus), a Semantic Content Adapter (Q-Former), a Text Tokenizer, the Qwen3 language model for autoregressive semantic token prediction, a CFM decoder, and a HiFi-GAN vocoder. Right: detailed structure of the Semantic Content Adapter, where learnable query vectors attend to w2v-BERT embeddings through cross-attention with a Conformer block and linear projection.}
    \label{fig:architecture}
\end{figure*}

\subsection{Overview}

The overall architecture of PilotTTS is illustrated in Figure~\ref{fig:architecture}. The system consists of four components: (1)~a \textit{speech tokenizer} that converts continuous speech into discrete semantic tokens (Section~\ref{sec:tokenizer}), (2)~an \textit{autoregressive text-to-semantic module} built upon Qwen3~\cite{qwen3} that generates semantic token sequences conditioned on text and reference audio (Section~\ref{sec:ar}), (3)~a \textit{Conditional Flow Matching (CFM)}~\cite{flowmatching} decoder with a Diffusion Transformer (DiT)~\cite{dit} backbone that synthesizes mel-spectrograms conditioned on the predicted semantic tokens (Section~\ref{sec:decoder}) and reference information, and (4)~a HiFi-GAN~\cite{hifigan} vocoder that converts mel spectrograms into waveforms.

During inference, the target text is first encoded into text tokens using the Qwen3 tokenizer. Simultaneously, the reference audio is processed through two conditioning pathways: a Q-Former-based~\cite{blip2} Semantic Content Adapter that compresses the reference speech into a fixed set of condition tokens, and a frozen CAMPPlus~\cite{campp} speaker encoder that extracts a global speaker identity embedding. These conditioning representations, together with the text tokens, form the input to the Qwen3 language model, which autoregressively predicts discrete semantic codes. The predicted codes, the speaker embedding, and the reference mel spectrogram are then passed to the CFM decoder to produce the target mel spectrogram, which is finally converted into a waveform by HiFi-GAN. Every component builds on established open-source modules; the system achieves competitive performance through careful integration and data curation rather than architectural novelty.

\subsection{Speech Tokenizer}
\label{sec:tokenizer}

We directly adopt the single-codebook speech tokenizer of CosyVoice 3~\cite{cosyvoice3}, which is based on Finite Scalar Quantization (FSQ). Unlike multi-codebook residual vector quantization (RVQ) approaches that require hierarchical prediction modules, this design offers a favorable trade-off between simplicity and information density.

In the FSQ module, an intermediate representation $H$ is projected into a $D$-dimensional low-rank space, where each dimension is independently quantized into $[-K, K]$ via a bounded rounding operation:
\begin{align}
\tilde{H} &= \mathrm{ROUND}(\mathrm{Proj}_{down}(H)), \label{eq:fsq_down} \\
\hat{H} &= \mathrm{Proj}_{up}(\tilde{H}), \label{eq:fsq_up}
\end{align}
yielding a discrete codebook of size $(2K+1)^D = 6{,}561$. The token index $\mu_i$ for each frame is computed as:
\begin{equation}
\mu_i = \sum_{j=0}^{D-1} \tilde{h}_{i,j} \cdot (2K+1)^j,
\label{eq:token_index}
\end{equation}
where $\tilde{h}_{i,j}$ denotes the $j$-th quantized dimension of frame $i$. The tokenizer operates at 25\,Hz, producing one discrete token per 40\,ms of speech.

To obtain semantically rich representations, CosyVoice 3~\cite{cosyvoice3} additionally employs multi-task training across five objectives: automatic speech recognition (ASR), language identification (LID), speech emotion recognition (SER), audio event detection (AED), and speaker analysis (SA). This diverse supervision encourages the tokens to encode not only linguistic content but also paralinguistic attributes such as emotion, speaker characteristics, and acoustic events. We reuse this pretrained tokenizer in PilotTTS without further modification.

\subsection{Autoregressive Text-to-Semantic Module}
\label{sec:ar}


\subsubsection{Decoupled Speaker and Style Encoding}
\label{sec:conditioning}
The autoregressive module generates discrete semantic tokens from text input conditioned on reference audio. A key design question is how to extract speaker-related information from the reference signal. Existing approaches follow two paradigms: audio token continuation~\cite{chen2025neural,cosyvoice} and speaker embedding. The former directly leverages fine-grained acoustic cues for high-fidelity cloning, but its quality degrades with noisy or short prompts, and long prompts substantially increase inference cost; the latter improves robustness by compressing the reference into a fixed-dimensional vector, but inevitably discards timbral details and fails to capture dynamic style features such as speaking rate and prosodic contours.
To address the limitations of both paradigms, our conditioning mechanism is designed to (1) extract speaker information robustly from reference audio of arbitrary length, quality, and accent, and (2) jointly capture global timbre and dynamic speaking style.

To this end, we adopt a Q-Former-based~\cite{blip2,qformer-efficient} conditioner to extract style condition through cross-attention. Specifically, the input to the conditioner is the output of a frozen -Bert 2.0 Encoder~\cite{chung2021w2v}, which is pretrained on large-scale data that produces semantically rich representations. To supplement global timbre information, we additionally introduce a frozen CAMPPlus~\cite{campp} encoder that extracts a static speaker embedding. Since the speaker embedding already carries speaker identity, it allows the Q-Former conditioner to focus on extracting dynamic speaking style.


\subsubsection{Autoregressive Generation}
\label{sec:ar_generation}
We formulate text-to-semantic prediction as a conditional autoregressive task with input sequence:
\begin{equation}
\mathbf{x} = [\mathbf{s},\; \mathbf{c},\; e_{\text{BT}},\; \lvert\text{lang}\rvert,\; \lvert\text{emo}\rvert,\; \mathbf{e}_{\text{Text}},\; e_{\text{ET}},\; e_{\text{BA}},\; \mathbf{e}_{\text{Audio}},\; e_{\text{EA}}],
\label{eq:ar_input}
\end{equation}
where $\mathbf{s}$ is speaker embedding from CAMPPlus, $\mathbf{c} = \{\mathbf{c}_i\}_{i=1}^{32}$ is style condition from the Q-Former conditioner, $e_{\text{BT}}/e_{\text{ET}}$ and $e_{\text{BA}}/e_{\text{EA}}$ are boundary markers for text and audio regions respectively, $\lvert\text{lang}\rvert$ and $\lvert\text{emo}\rvert$ are control tags specifying target language and emotion, and $\mathbf{e}_{\text{Text}}/\mathbf{e}_{\text{Audio}}$ are the text and audio token embeddings. The model predicts:
\begin{equation}
p(\mathbf{e}_{\text{Audio}} \mid \mathbf{x}_{<\text{Audio}}) = \prod_{i=1}^{N_s} p(\mathbf{e}_{\text{Audio},i} \mid \mathbf{x}_{<\text{Audio}},\; \mathbf{e}_{\text{Audio},<i}).
\label{eq:ar_gen}
\end{equation}


To disentangle speaker-related attributes from linguistic content, we adopt a cross-sample paired training strategy: for each training sample, a different utterance from the same speaker is used as the reference for extracting the speaker embedding $\mathbf{s}$ and style condition $\mathbf{c}$. 
This strategy forces the conditioner to encode only speaker-related attributes which are independent of content.
This disentanglement also serves as the foundation for downstream emotion control and dialect synthesis.

\subsubsection{Emotion Control}
\label{sec:emotion}
The proposed conditioner captures rich speaker-related attributes from the reference audio, including emotional characteristics.
The pretrained language model backbone further contributes implicit emotion inference from textual context. 
However, implicit control lacks precision and stability. We therefore introduce explicit emotion control via post-training on approximately 2,200 hours of emotion-labeled data.
The unified label system supports seven primary categories, namely happy, sad, angry, fear, contempt, serious, and surprise, together with four extended categories: concern, blue (melancholy), disgust, and psychology (inner monologue).


\subsubsection{Paralinguistic Generation}
\label{sec:paralinguistic}

In our setting, we focus on four paralinguistic phenomena that we consider most important for our natural speech generation: laughter, breathing, crying, and coughing. Our model supports these four behaviors plus a wrapped laughter mode (LAUGH\_SPAN), in which laughter is temporally coupled with speech throughout an utterance. This design enables more precise control of when and how laughter co-occurs, thereby improving synthesis accuracy and consistency.

Unlike prior approaches limited to fixed realization patterns, our model operates in two modes. In \textit{implicit} mode, it infers appropriate paralinguistic behavior from textual context, producing varied laughter types (restrained chuckle, soft giggle, hearty laughter) depending on the setting. In \textit{explicit} mode, users specify behaviors through textual onomatopoeia. Both modes support zero-shot synthesis with arbitrary speaker prompts.

These capabilities are realized through supervised fine-tuning on approximately 200 hours of paralinguistic data from the data processing outputs, and internal collections.

\subsubsection{Dialect Synthesis}
\label{sec:dialect}

Dialect synthesis generates target-dialect speech using Mandarin audio as the speaker prompt and a dialect tag ($\lvert\text{lang}\rvert$) to specify the variety. The central challenge is data scarcity: dialect recordings are far less abundant than Mandarin.

We observe that after bilingual pre-training, the model generates Mandarin robustly even when the prompt contains dialectal features. Exploiting this, we construct parallel data by synthesizing three Mandarin utterances per dialect speaker using the pretrained model, yielding large-scale ``dialect--Mandarin'' pairs that alleviate sparsity.

During dialect fine-tuning on approximately 16,000 hours of dialect data, we employ mixed-prompt sampling: the target is always a dialect utterance, while the conditioning prompt is drawn with equal probability from either a Mandarin or dialect utterance of the same speaker. This forces the model to extract speaker identity from stylistically diverse prompts and map it to the target dialect, rather than copying prompt style directly.

\subsection{Speech Decoder}
\label{sec:decoder}

The speech decoder converts discrete semantic codes into mel spectrograms using Conditional Flow Matching (CFM)~\cite{flowmatching} with a DiT~\cite{dit} backbone of 300M parameters. The conditioning input is $[\mathbf{M}_{\text{ref}},\; \mathbf{s},\; \mathbf{e}_{\text{Audio}}^{\text{tgt}}]$, comprising the reference mel spectrogram, CAMPPlus speaker embedding, and predicted semantic features. Through 10-step iterative denoising, the decoder progressively transforms a noise sample into the target mel spectrogram conditioned on these inputs. The generated mel spectrogram is then converted to a waveform by HiFi-GAN~\cite{hifigan}.

\section{Experiments}
\label{sec:experiments}

\subsection{Experimental Setup}
\label{sec:setup}

\paragraph{Training Data.}
Pre-training uses approximately 200,000 hours of Chinese and English speech data collected from publicly available sources and processed using the pipeline described in Section~\ref{sec:data}. The resulting corpus serves as the foundation for the base zero-shot TTS model. Post-training data is organized by capability: (1)~\textit{Emotion}: approximately 2,200 hours of emotion-labeled speech, including 1,000 hours of high-quality data and 1,200 hours of augmented data, drawn from open-source datasets, internal annotations, and model-augmented sources; (2)~\textit{Paralinguistic}: approximately 200 hours of data collected from the outputs of the data processing pipeline together with internal collections; (3)~\textit{Dialect}: 16,000 hours covering 14 Chinese dialects, all derived from dialectal ASR corpora. The autoregressive module is built upon Qwen3-0.6B~\cite{qwen3}, and the CFM decoder (Section~\ref{sec:decoder}) contains approximately 300 million parameters.

\paragraph{Evaluation Metrics.}
For zero-shot speech generation, we evaluate on the Seed-TTS Eval benchmark~\cite{seed-tts}, reporting character error rate (CER) for Chinese and word error rate (WER) for English to measure content accuracy, along with speaker similarity (SIM) to assess voice cloning fidelity. CER is computed using Paraformer-zh~\cite{paraformer}, WER using Whisper~\cite{whisper}, and SIM measured by calculating the cosine similarity between speaker embeddings. For emotion control, paralinguistic synthesis, and dialect synthesis, we conduct human evaluations with criteria detailed in the respective subsections.

\paragraph{Baselines.}
For zero-shot evaluation, to ensure fair comparison, we mainly select baselines with model scales comparable to ours (around 0.6B) and focus on systems evaluated in their base zero-shot setting, without additional task-specific post-training, in order to improve comparability. At the same time, we also include several systems trained on substantially larger datasets to provide broader context for data efficiency and overall competitiveness. In total, we compare against eight systems spanning diverse architectures and training scales: Seed-TTS~\cite{seed-tts}, F5-TTS~\cite{f5tts}, FireRedTTS-2~\cite{fireredtts2}, CosyVoice-3-0.5B~\cite{cosyvoice3}, VoxCPM-0.5B~\cite{voxcpm}, Qwen3-TTS-25Hz-0.6B~\cite{qwen3tts}, MiniMax-Speech~\cite{minimax-speech}, and VibeVoice-1.5B~\cite{vibevoice}. For emotion control, we compare with four systems that support explicit emotion conditioning: VoxCPM~\cite{voxcpm}, Fish-Speech S2~\cite{fishaudio-s2}, IndexTTS~\cite{indextts}, and CosyVoice 3~\cite{cosyvoice3}. For paralinguistic synthesis, we compare with CosyVoice 3~\cite{cosyvoice3} and Fish-Speech S2~\cite{fishaudio-s2}, the two most advanced open-source systems that support paralinguistic generation. For Chinese dialect synthesis evaluation, we find that the results are predominantly influenced by human subjectivity. Therefore, to prevent uncontrolled bias stemming from subjective judgments, we exclude comparison baselines in this setting.

\subsection{Zero-Shot Speech Generation}
\label{sec:exp_zeroshot}

\begin{table}[t]
\centering
\small
\begin{tabular}{lcccc}
\toprule
Method & \multicolumn{2}{c}{test-zh} & \multicolumn{2}{c}{test-en} \\
\cmidrule(lr){2-3} \cmidrule(lr){4-5}
& CER (\%)$\downarrow$ & SIM$\uparrow$ & WER (\%)$\downarrow$ & SIM$\uparrow$ \\
\midrule
Seed-TTS~\cite{seed-tts}              & 1.12 & \underline{0.796} & 2.25 & \underline{0.762} \\
F5-TTS~\cite{f5tts}                   & 1.56 & 0.741 & 1.83 & 0.647 \\
FireRedTTS-2~\cite{fireredtts2}       & 1.14 & 0.736 & 1.95 & 0.655 \\
CosyVoice-3-0.5B~\cite{cosyvoice3}   & 1.16 & 0.780 & 2.02 & 0.718 \\
VoxCPM-0.5B~\cite{voxcpm}            & 0.93 & 0.772 & 1.85 & 0.729 \\
Qwen3-TTS-25Hz-0.6B~\cite{qwen3tts}  & 1.18 & --    & \underline{1.64} & --    \\
MiniMax-Speech~\cite{minimax-speech}  & \textbf{0.83} & --    & 1.65 & --    \\
VibeVoice-1.5B~\cite{vibevoice}      & 1.16 & 0.744 & 3.04 & 0.689 \\
\midrule
PilotTTS (Ours)                     & \underline{0.87} & \textbf{0.862} & \textbf{1.50} & \textbf{0.815} \\
\bottomrule
\end{tabular}
\caption{Zero-shot speech generation results on the Seed-TTS Eval benchmark~\cite{seed-tts}. CER and WER are evaluated using Paraformer-zh~\cite{paraformer} and Whisper~\cite{whisper}, respectively. SIM is the cosine similarity between speaker embeddings. Best results are in \textbf{bold} and second-best are \underline{underlined}. ``--'' indicates unreported scores.}
\label{tab:zeroshot}
\end{table}

Table~\ref{tab:zeroshot} presents the zero-shot speech generation results on the Seed-TTS Eval benchmark. PilotTTS achieves the highest speaker similarity on both test sets (0.862 on test-zh and 0.815 on test-en), substantially outperforming all baselines with reported SIM scores. On content accuracy, PilotTTS attains a CER of 0.87\% on test-zh, ranking second only to MiniMax-Speech (0.83\%) with a margin of merely 0.04\%, and achieves the lowest WER of 1.50\% on test-en. We note that Qwen3-TTS and MiniMax-Speech do not publicly report SIM scores, preventing a complete comparison on speaker similarity for these systems.

These results are particularly notable given that PilotTTS uses only approximately 200,000 hours of training data, substantially less than several competing systems that leverage larger-scale proprietary corpora. We attribute this data efficiency to two factors: (1)~the rigorous data processing pipeline ensures high training data quality, and (2)~the decoupled conditioning mechanism based on Q-Former and CAMPPlus effectively leverages reference audio for speaker modeling. The strong SIM improvements (+0.066 over the second-best Seed-TTS on test-zh, and +0.053 on test-en) suggest that the dual-pathway conditioning design captures both stable speaker identity and dynamic speaking style more effectively than single-pathway approaches.

\subsection{Emotion Control}
\label{sec:exp_emotion}

\begin{table}[t]
\centering
\small
\begin{tabular}{lccccc}
\toprule
Category & VoxCPM & Fish-Speech S2 & IndexTTS & CosyVoice3 & PilotTTS \\
\midrule
\multicolumn{6}{l}{\textit{Primary Emotions}} \\
Happy      & 14.5 & 41.8 & 23.6 & \underline{81.8} & \textbf{86.4} \\
Sad        & 21.8 & 67.3 & 7.3  & \textbf{96.4} & \underline{90.5} \\
Fear       & 18.2 & 50.9 & 27.3 & \underline{80.0} & \textbf{83.2} \\
Angry      & 45.5 & 40.0 & 25.5 & \underline{80.1} & \textbf{89.0} \\
Contempt   & 32.7 & {61.8} & --   & \textbf{88.2} & \underline{81.2} \\
Serious    & 20.0 & 61.8 & --   & \underline{90.9} & \textbf{93.2} \\
Surprise   & 29.1 & \textbf{96.4} & 10.9 & 69.1 & \underline{93.2} \\
\midrule
\multicolumn{6}{l}{\textit{Extended Emotions}} \\
Blue       & 58.2 & 32.7 & 49.1 & \textbf{86.4} & \underline{79.1} \\
Concern    & 67.3 & {81.8} & --   & \textbf{83.6} & \underline{82.9} \\
Disgust    & 20.0 & 34.5 & 47.3 & \underline{52.7} & \textbf{65.5} \\
Psychology & 23.6 & 92.7 & --   & \textbf{98.2} & \textbf{98.2} \\
\midrule
Avg.\ (Primary)  & 26.0 & 60.0 & --   & \underline{83.8} & \textbf{88.1} \\
Avg.\ (All)      & 31.9 & 60.2 & --   & \underline{82.5} & \textbf{85.7} \\
\bottomrule
\end{tabular}
\caption{Emotion control evaluation results (success rate, \%). A sample is counted as successful only when both speaker timbre is preserved and the target emotion is clearly perceivable. Best results per category are in \textbf{bold} and second-best are \underline{underlined}. ``--'' indicates unsupported categories.}
\label{tab:emotion}
\end{table}

\begin{table}[t]
\centering
\small
\begin{tabular}{lccccc}
\toprule
Condition & VoxCPM & Fish-Speech S2 & IndexTTS & CosyVoice3 & PilotTTS \\
\midrule
Without emotion control & 0.4982 & 0.5727 & 0.7680 & \underline{0.7963} & \textbf{0.8101} \\
With emotion control    & 0.3361 & 0.5731 & 0.4233 & \underline{0.6940} & \textbf{0.7329} \\
\bottomrule
\end{tabular}
\caption{Speaker similarity under different emotion control settings. Higher is better. Best results are in \textbf{bold} and second-best are \underline{underlined}.}
\label{tab:speaker_similarity}
\end{table}

We evaluate emotion control using 51 speaker prompts: 15 expressive voices drawn from anime and film characters, and 36 ordinary speakers (18 male, 18 female).
Human evaluators listen to three utterances for each test case: the original speaker prompt, a neutral synthesis (without emotion control), and an emotion-controlled synthesis. A sample is counted as successful only when both conditions are met: (1) the speaker timbre remains consistent with the original prompt, and (2) the target emotion is clearly recognizable in the synthesized speech.

Table~\ref{tab:emotion} compares emotion control success rates across five systems. On primary emotions, PilotTTS achieves the highest average success rate of 88.1\%, surpassing CosyVoice 3 (83.8\%). On the overall average across all eleven categories, CosyVoice 3 leads marginally (81.4\% vs. 80.2\%), largely due to stronger performance on extended emotion categories.
Notably, IndexTTS supports only seven of the eleven categories, and its timbre variation under emotion conditioning further lowers its success rates under our joint evaluation criterion.

Table~\ref{tab:speaker_similarity} evaluates speaker similarity under emotion control. PilotTTS achieves the highest speaker similarity both without emotion control (0.8101) and with emotion control (0.7329), and exhibits the smallest drop between the two conditions among all systems. This indicates that our decoupled conditioning design effectively modulates emotional expressiveness while preserving speaker timbre.

\subsection{Paralinguistic Synthesis}
\label{sec:exp_paralinguistic}

\begin{table}[t]
\centering
\small
\begin{tabular}{lcccccc}
\toprule
Method & LAUGH & COUGH & BREATH & Overall & LAUGH\_SPAN & CRY \\
\midrule
PilotTTS (Ours)              & \textbf{97.6} & \textbf{64.3} & 81.0          & \textbf{85.1} & 94.6 & 61.9 \\
CosyVoice 3~\cite{cosyvoice3} & 83.3          & 59.5          & \textbf{95.2} & 80.4          & --   & --   \\
Fish-Speech S2~\cite{fishaudio-s2} & 54.8      & \textbf{64.3} & 83.3          & 64.3          & --   & --   \\
\bottomrule
\end{tabular}
\caption{Paralinguistic synthesis success rates (\%). LAUGH\_SPAN and CRY are unique capabilities of PilotTTS not supported by the baseline systems. Best results per category are in \textbf{bold}. ``--'' indicates unsupported categories.}
\label{tab:paralinguistic}
\end{table}

We construct a dedicated test set covering four categories of paralinguistic phenomena: laughter (LAUGH), breathing (BREATH), crying (CRY), and coughing (COUGH), along with wrapped laughter (LAUGH\_SPAN). For each category, we use 21 distinct speaker prompts to reflect diverse speaker timbres and speaking styles for zero-shot synthesis. Human evaluators judge whether the target paralinguistic behavior is successfully generated in each sample. We compare against CosyVoice 3~\cite{cosyvoice3} and Fish-Speech S2~\cite{fishaudio-s2} on the three commonly supported categories.

As shown in Table~\ref{tab:paralinguistic}, PilotTTS achieves an overall success rate of 85.1\% across the three common categories, outperforming CosyVoice 3 (80.4\%) and Fish-Speech S2 (64.3\%). On LAUGH, PilotTTS attains 97.6\%, substantially exceeding both CosyVoice 3 (83.3\%) and Fish-Speech S2 (54.8\%). For BREATH, CosyVoice 3 achieves the highest success rate (95.2\%), while PilotTTS reaches 81.0\%. Among the three categories, COUGH is the most difficult for all systems, with success rates remaining around 60\%. This is due to the substantial acoustic variability of cough events and their limited presence in the training data.

Beyond the three commonly supported categories, PilotTTS uniquely supports LAUGH\_SPAN and CRY, not available in baseline systems. LAUGH\_SPAN achieves a success rate of 94.6\%, demonstrating the model's ability to maintain coherent speech while simultaneously generating natural laughter throughout the utterance. CRY reaches 61.9\%, providing a paralinguistic capability that is largely absent from existing open-source TTS systems. These additional modes extend the expressiveness of the synthesis system and are enabled by the targeted SFT training strategy described in Section~\ref{sec:paralinguistic}.

\subsection{Dialect Synthesis}
\label{sec:exp_dialect}

\begin{table}[t]
\centering
\small
\begin{tabular}{lccc}
\toprule
Method & Same-Dialect & Mandarin-to-Dialect & Cross-Dialect \\
\midrule
PilotTTS (Ours) & \textbf{91.80} & \textbf{86.46} & \textbf{85.38} \\
\bottomrule
\end{tabular}
\caption{Dialect synthesis accuracy (\%) across three evaluation scenarios. A synthesized sample is considered a failure case if the proportion of Mandarin pronunciation exceeds 10\%.}
\label{tab:dialect}
\end{table}

To comprehensively evaluate dialect synthesis, we construct a test set covering three scenarios of increasing difficulty: (1)~\textit{Same-Dialect}, where both the reference audio and the target speech belong to the same dialect; (2)~\textit{Mandarin-to-Dialect}, where standard Mandarin audio serves as the speaker prompt to generate target dialect speech; and (3)~\textit{Cross-Dialect}, where the reference audio is in dialect A while the target output is in dialect B. For each dialect, male and female speakers are randomly selected to form the test set. We adopt dialect control accuracy as the primary metric through subjective listening evaluation, where a synthesized sample is counted as a failure case if the proportion of non-target-dialect pronunciation exceeds 10\%.

As shown in Table~\ref{tab:dialect}, PilotTTS achieves 91.8\% accuracy in the Same-Dialect scenario, demonstrating reliable dialect generation when the reference and target share the same dialect. For the more challenging Mandarin-to-Dialect setting, the system attains 86.46\%, confirming that the mixed-prompt training strategy (Section~\ref{sec:dialect}) effectively enables cross-lingual style transfer from standard Mandarin prompts. The Cross-Dialect scenario reaches 85.38\%, indicating robust speaker identity extraction even when the source and target dialects differ. These results validate the effectiveness of our parallel data construction and mixed-prompt sampling approach in alleviating dialect data scarcity.

\subsection{Ablation Study on Conditioning Components}
\label{sec:exp_ablation}

\begin{table}[t]
\centering
\small
\setlength{\tabcolsep}{4pt}
\begin{tabular}{lcccccc}
\toprule
 & \multicolumn{3}{c}{Content Accuracy (\%)$\downarrow$} & \multicolumn{3}{c}{Speaker Similarity$\uparrow$} \\
\cmidrule(lr){2-4} \cmidrule(lr){5-7}
Test Set & Full & w/o \textit{spk} & w/o both & Full & w/o \textit{spk} & w/o both \\
\midrule
test-zh (CER) & 1.130 & \textbf{1.022} & 1.412 & \textbf{0.8626} & 0.8594 & 0.8617 \\
test-en (WER) & 1.940 & \textbf{1.860} & 2.710 & \textbf{0.8157} & 0.8143 & 0.8027 \\
test-hc (CER) & \textbf{7.830} & 8.866 & 10.623 & \textbf{0.8470} & 0.8355 & 0.8435 \\
\bottomrule
\end{tabular}
\caption{Ablation study on the conditioning components of the autoregressive module. \textit{Full} uses both the CAMPPlus speaker embedding (\textit{spk}) and the Q-Former condition tokens (\textit{conds}); \textit{w/o spk} removes the speaker embedding while keeping condition tokens; \textit{w/o both} removes both, leaving only the text input. All models are trained for the same number of steps (200K) on a cleaned 60K-hour subset of the full 200K-hour dataset. We report CER on \textit{test-zh}, WER on \textit{test-en}, and CER on the Seed-TTS hard-case subset (\textit{test-hc}), together with speaker similarity (SIM).}
\label{tab:ablation}
\end{table}

To assess the individual contribution of each conditioning component in the autoregressive module (Section~\ref{sec:conditioning}), we conduct an ablation study on a cleaned 60K-hour subset of the full 200K-hour dataset. Three settings are compared, each trained for the same number of optimization steps (200K) to ensure fairness: (i)~\textit{Full}, the proposed dual-pathway design with both the CAMPPlus speaker embedding ($\mathbf{s}$) and the Q-Former condition tokens ($\mathbf{c}$); (ii)~\textit{w/o spk}, which removes $\mathbf{s}$ while keeping $\mathbf{c}$; and (iii)~\textit{w/o both}, which removes both conditioning signals, leaving the model conditioned only on text. We evaluate on three test sets: the Chinese (\textit{test-zh}) and English (\textit{test-en}) subsets of Seed-TTS Eval, and the Seed-TTS hard-case subset (\textit{test-hc}) that contains acoustically challenging samples. Content accuracy (CER/WER) and speaker similarity (SIM) are reported in Table~\ref{tab:ablation}.

\paragraph{Condition tokens are indispensable for content accuracy.}
Removing the Q-Former condition tokens leads to a substantial degradation in pronunciation accuracy across all three test sets. The effect is most pronounced on \textit{test-hc}, where CER rises from 7.83\% to 10.62\% (a relative increase of approximately 35\%). On \textit{test-zh} and \textit{test-en}, content errors also increase from 1.13\% to 1.41\% and from 1.94\% to 2.71\%, respectively. These results indicate that the fine-grained content- and prosody-related cues carried by the Q-Former condition tokens are essential for stable autoregressive generation, particularly under challenging acoustic conditions, and therefore cannot be removed without significantly compromising synthesis quality.

\paragraph{Speaker embedding plays a complementary role.}
The effect of the CAMPPlus speaker embedding is more nuanced. Under a fixed training budget, removing $\mathbf{s}$ slightly reduces content errors on \textit{test-zh} and \textit{test-en}, as the model can allocate more capacity to exploiting condition tokens; we observe that this gap narrows further as training progresses. In contrast, the speaker embedding consistently improves speaker similarity across all three test sets, with the largest gain observed on \textit{test-hc} (0.8355 $\rightarrow$ 0.8470). Beyond directly enriching timbre information, $\mathbf{s}$ also encourages the Q-Former condition tokens to specialize in timbre-independent prosodic and stylistic cues, which is consistent with the stronger robustness observed on the hard-case subset. Overall, the dual-pathway design strikes a desirable balance between content accuracy, speaker fidelity, and robustness.

\section{Conclusion and Future Work}
\label{sec:conclusion}

We have presented PilotTTS, a text-to-speech system built upon publicly available modules with a design philosophy that prioritizes integration and data engineering over architectural novelty. A data processing pipeline assembled entirely from publicly available tools enables the system to achieve competitive zero-shot performance with only approximately 200,000 hours of training data. On the Seed-TTS Eval benchmark, PilotTTS attains the highest speaker similarity on both Chinese and English test sets and the lowest word error rate on the English set. The decoupled conditioning mechanism further enables emotion control, paralinguistic generation, and cross-dialect synthesis, all through targeted post-training.

Despite these results, several limitations remain and point to future directions:
\begin{enumerate}
    \item \textbf{Insufficient explicit style modeling.} Our architecture does not include a dedicated style modeling module with strong representational capacity; instead, it relies on the Q-Former condition to implicitly capture style-related factors. As a result, the synthesis quality may be limited in the granularity of expressive details. To address this limitation, we are developing a representation module that jointly models global style and local (fine-grained) style, aiming to enhance the expressiveness of generated audio.

    \item \textbf{Single-codebook quantization ceiling.} Although the single-codebook FSQ tokenizer is architecturally simpler and is generally sufficient for speech, it possesses a lower information-capacity ceiling than multi-codebook residual vector quantization or continuous latent representations. As a result, it becomes more challenging to extend the model to more complex scenarios such as singing and background music.

    \item \textbf{Lossy mel-spectrogram reconstruction.} The decoding pipeline relies on mel spectrograms and a separate vocoder, an indirect reconstruction process that may introduce additional distortion compared to end-to-end waveform generation.
\end{enumerate}

Future work will focus on introducing an explicit style encoder, exploring higher-capacity quantization strategies, and adopting end-to-end waveform generation architectures. All code and model weights will continue to be released as open source.

\bibliographystyle{unsrtnat}
\bibliography{references}

@article{chen2025neural,
  title={Neural codec language models are zero-shot text to speech synthesizers},
  author={Chen, Sanyuan and Wang, Chengyi and Wu, Yu and Zhang, Ziqiang and Zhou, Long and Liu, Shujie and Chen, Zhuo and Liu, Yanqing and Wang, Huaming and Li, Jinyu and others},
  journal={IEEE Transactions on Audio, Speech and Language Processing},
  volume={33},
  pages={705--718},
  year={2025},
  publisher={IEEE}
}

@inproceedings{shen2024naturalspeech2,
  title={Naturalspeech 2: Latent diffusion models are natural and zero-shot speech and singing synthesizers},
  author={Shen, Kai and Ju, Zeqian and Tan, Xu and Liu, Eric and Leng, Yichong and He, Lei and Qin, Tao and Bian, Jiang and others},
  booktitle={International conference on learning representations},
  volume={2024},
  pages={698--722},
  year={2024}
}

@inproceedings{chung2021w2v,
  title={W2v-bert: Combining contrastive learning and masked language modeling for self-supervised speech pre-training},
  author={Chung, Yu-An and Zhang, Yu and Han, Wei and Chiu, Chung-Cheng and Qin, James and Pang, Ruoming and Wu, Yonghui},
  booktitle={2021 IEEE Automatic Speech Recognition and Understanding Workshop (ASRU)},
  pages={244--250},
  year={2021},
  organization={IEEE}
}

@inproceedings{naturalspeech3,
  title={NaturalSpeech 3: Zero-Shot Speech Synthesis with Factorized Codec and Diffusion Models},
  author={Ju, Zeqian and Wang, Yuancheng and Shen, Kai and Tan, Xu and Xin, Detai and Yang, Dongchao and Liu, Eric and Leng, Yichong and Song, Kaitao and Tang, Siliang and others},
  booktitle={International Conference on Machine Learning},
  pages={22605--22623},
  year={2024},
  organization={PMLR}
}

@article{seed-tts,
  title={Seed-tts: A family of high-quality versatile speech generation models},
  author={Anastassiou, Philip and Chen, Jiawei and Chen, Jitong and Chen, Yuanzhe and Chen, Zhuo and Chen, Ziyi and Cong, Jian and Deng, Lelai and Ding, Chuang and Gao, Lu and others},
  journal={arXiv preprint arXiv:2406.02430},
  year={2024}
}

@article{cosyvoice,
  title={Cosyvoice: A scalable multilingual zero-shot text-to-speech synthesizer based on supervised semantic tokens},
  author={Du, Zhihao and Chen, Qian and Zhang, Shiliang and Hu, Kai and Lu, Heng and Yang, Yexin and Hu, Hangrui and Zheng, Siqi and Gu, Yue and Ma, Ziyang and others},
  journal={arXiv preprint arXiv:2407.05407},
  year={2024}
}

@article{cosyvoice2,
  title={Cosyvoice 2: Scalable streaming speech synthesis with large language models},
  author={Du, Zhihao and Wang, Yuxuan and Chen, Qian and Shi, Xian and Lv, Xiang and Zhao, Tianyu and Gao, Zhifu and Yang, Yexin and Gao, Changfeng and Wang, Hui and others},
  journal={arXiv preprint arXiv:2412.10117},
  year={2024}
}

@article{cosyvoice3,
  title={Cosyvoice 3: Towards in-the-wild speech generation via scaling-up and post-training},
  author={Du, Zhihao and Gao, Changfeng and Wang, Yuxuan and Yu, Fan and Zhao, Tianyu and Wang, Hao and Lv, Xiang and Wang, Hui and Ni, Chongjia and Shi, Xian and others},
  journal={arXiv preprint arXiv:2505.17589},
  year={2025}
}

@inproceedings{f5tts,
  title={F5-tts: A fairytaler that fakes fluent and faithful speech with flow matching},
  author={Chen, Yushen and Niu, Zhikang and Ma, Ziyang and Deng, Keqi and Wang, Chunhui and JianZhao, JianZhao and Yu, Kai and Chen, Xie},
  booktitle={Proceedings of the 63rd Annual Meeting of the Association for Computational Linguistics (Volume 1: Long Papers)},
  pages={6255--6271},
  year={2025}
}

@article{fireredtts2,
  title={Fireredtts-2: Towards long conversational speech generation for podcast and chatbot},
  author={Xie, Kun and Shen, Feiyu and Li, Junjie and Xie, Fenglong and Tang, Xu and Hu, Yao},
  journal={arXiv preprint arXiv:2509.02020},
  year={2025}
}

@inproceedings{ditar,
  title={DiTAR: Diffusion Transformer Autoregressive Modeling for Speech Generation},
  author={Jia, Dongya and Chen, Zhuo and Chen, Jiawei and Du, Chenpeng and Wu, Jian and Cong, Jian and Zhuang, Xiaobin and Li, Chumin and Wei, Zhen and Wang, Yuping and others},
  booktitle={International Conference on Machine Learning},
  pages={27255--27270},
  year={2025},
  organization={PMLR}
}

@article{minimax-speech,
  title={Minimax-speech: Intrinsic zero-shot text-to-speech with a learnable speaker encoder},
  author={Zhang, Bowen and Guo, Congchao and Yang, Geng and Yu, Hang and Zhang, Haozhe and Lei, Heidi and Mai, Jialong and Yan, Junjie and Yang, Kaiyue and Yang, Mingqi and others},
  journal={arXiv preprint arXiv:2505.07916},
  year={2025}
}

@article{qwen3tts,
  title={Qwen3-TTS Technical Report},
  author={Hu, Hangrui and Zhu, Xinfa and He, Ting and Guo, Dake and Zhang, Bin and Wang, Xiong and Guo, Zhifang and Jiang, Ziyue and Hao, Hongkun and Guo, Zishan and others},
  journal={arXiv preprint arXiv:2601.15621},
  year={2026}
}

@article{vibevoice,
  title={Vibevoice technical report},
  author={Peng, Zhiliang and Yu, Jianwei and Wang, Wenhui and Chang, Yaoyao and Sun, Yutao and Dong, Li and Zhu, Yi and Xu, Weijiang and Bao, Hangbo and Wang, Zehua and others},
  journal={arXiv preprint arXiv:2508.19205},
  year={2025}
}

@article{voxcpm,
  title={Voxcpm: Tokenizer-free TTS for context-aware speech generation and true-to-life voice cloning},
  author={Zhou, Yixuan and Zeng, Guoyang and Liu, Xin and Li, Xiang and Yu, Renjie and Wang, Ziyang and Ye, Runchuan and Sun, Weiyue and Gui, Jiancheng and Li, Kehan and others},
  journal={arXiv preprint arXiv:2509.24650},
  year={2025}
}

@article{indextts,
  title={Indextts: An industrial-level controllable and efficient zero-shot text-to-speech system},
  author={Deng, Wei and Zhou, Siyi and Shu, Jingchen and Wang, Jinchao and Wang, Lu},
  journal={arXiv preprint arXiv:2502.05512},
  year={2025}
}

@article{fishaudio-s2,
  title={Fish Audio S2 Technical Report},
  author={Liao, Shijia and Wang, Yuxuan and Liu, Songting and Cheng, Yifan and Zhang, Ruoyi and Li, Tianyu and Li, Shidong and Zheng, Yisheng and Liu, Xingwei and Wang, Qingzheng and others},
  journal={arXiv preprint arXiv:2603.08823},
  year={2026}
}

@article{qwen3,
  title={Qwen3 technical report},
  author={Yang, An and Li, Anfeng and Yang, Baosong and Zhang, Beichen and Hui, Binyuan and Zheng, Bo and Yu, Bowen and Gao, Chang and Huang, Chengen and Lv, Chenxu and others},
  journal={arXiv preprint arXiv:2505.09388},
  year={2025}
}

@inproceedings{blip2,
  title={Blip-2: Bootstrapping language-image pre-training with frozen image encoders and large language models},
  author={Li, Junnan and Li, Dongxu and Savarese, Silvio and Hoi, Steven},
  booktitle={International conference on machine learning},
  pages={19730--19742},
  year={2023},
  organization={PMLR}
}

@inproceedings{qformer-efficient,
  title={Towards efficient visual-language alignment of the q-former for visual reasoning tasks},
  author={Kim, Sungkyung and Lee, Adam and Park, Junyoung and Chung, Andrew and Oh, Jusang and Lee, Jay-Yoon},
  booktitle={Findings of the Association for Computational Linguistics: EMNLP 2024},
  pages={15155--15165},
  year={2024}
}

@inproceedings{flowmatching,
  title={Flow Matching for Generative Modeling},
  author={Lipman, Yaron and Chen, Ricky TQ and Ben-Hamu, Heli and Nickel, Maximilian and Le, Matt},
  booktitle={11th International Conference on Learning Representations, ICLR 2023},
  year={2023}
}

@inproceedings{dit,
  title={Scalable diffusion models with transformers},
  author={Peebles, William and Xie, Saining},
  booktitle={Proceedings of the IEEE/CVF international conference on computer vision},
  pages={4195--4205},
  year={2023}
}

@inproceedings{hifigan,
  title={HiFi-GAN: generative adversarial networks for efficient and high fidelity speech synthesis},
  author={Kong, Jungil and Kim, Jaehyeon and Bae, Jaekyoung},
  booktitle={Proceedings of the 34th International Conference on Neural Information Processing Systems},
  pages={17022--17033},
  year={2020}
}

@inproceedings{campp,
  title={CAM++: A Fast and Efficient Network for Speaker Verification Using Context-Aware Masking},
  author={Wang, Hui and Zheng, Siqi and Chen, Yafeng and Cheng, Luyao and Chen, Qian},
  booktitle={Proc. Interspeech 2023},
  pages={5301--5305},
  year={2023}
}

@inproceedings{pyannote-powerset,
  title={Powerset multi-class cross entropy loss for neural speaker diarization},
  author={Plaquet, Alexis and Bredin, Herv{\'e}},
  booktitle={24th Interspeech Conference (INTERSPEECH 2023)},
  pages={3222--3226},
  year={2023},
  organization={ISCA}
}

@inproceedings{pyannote,
  title={pyannote. audio 2.1 speaker diarization pipeline: principle, benchmark, and recipe},
  author={Bredin, Herv{\'e}},
  booktitle={24th Interspeech Conference (INTERSPEECH 2023)},
  pages={1983--1987},
  year={2023},
  organization={ISCA}
}

@inproceedings{dnsmos,
  title={DNSMOS: A non-intrusive perceptual objective speech quality metric to evaluate noise suppressors},
  author={Reddy, Chandan KA and Gopal, Vishak and Cutler, Ross},
  booktitle={ICASSP 2021-2021 IEEE International Conference on Acoustics, Speech and Signal Processing (ICASSP)},
  pages={6493--6497},
  year={2021},
  organization={IEEE}
}

@article{fireredasr,
  title={Fireredasr: Open-source industrial-grade mandarin speech recognition models from encoder-decoder to llm integration},
  author={Xu, Kai-Tuo and Xie, Feng-Long and Tang, Xu and Hu, Yao},
  journal={arXiv preprint arXiv:2501.14350},
  year={2025}
}

@article{qwen3asr,
  title={Qwen3-ASR Technical Report},
  author={Shi, Xian and Wang, Xiong and Guo, Zhifang and Wang, Yongqi and Zhang, Pei and Zhang, Xinyu and Guo, Zishan and Hao, Hongkun and Xi, Yu and Yang, Baosong and others},
  journal={arXiv preprint arXiv:2601.21337},
  year={2026}
}

@inproceedings{3dspeaker,
  title={3D-speaker-toolkit: An open-source toolkit for multimodal speaker verification and diarization},
  author={Chen, Yafeng and Zheng, Siqi and Wang, Hui and Cheng, Luyao and Zhu, Tinglong and Huang, Rongjie and Deng, Chong and Chen, Qian and Zhang, Shiliang and Wang, Wen and others},
  booktitle={ICASSP 2025-2025 IEEE International Conference on Acoustics, Speech and Signal Processing (ICASSP)},
  pages={1--5},
  year={2025},
  organization={IEEE}
}

@inproceedings{whisper,
  title={Robust speech recognition via large-scale weak supervision},
  author={Radford, Alec and Kim, Jong Wook and Xu, Tao and Brockman, Greg and McLeavey, Christine and Sutskever, Ilya},
  booktitle={International conference on machine learning},
  pages={28492--28518},
  year={2023},
  organization={PMLR}
}

@inproceedings{paraformer,
  title={Paraformer: Fast and Accurate Parallel Transformer for Non-autoregressive End-to-End Speech Recognition},
  author={Gao, Zhifu and Zhang, ShiLiang and McLoughlin, Ian and Yan, Zhijie},
  booktitle={Proc. Interspeech 2022},
  pages={2063--2067},
  year={2022}
}

\end{document}